\documentclass[a4,12pt]{article}
\usepackage{graphicx}
\newcommand{\nc}{\newcommand}
\nc{\bg}{B. Grzadkowski}
\nc{\non}{\nonumber}
\def\dps{\displaystyle}
\def\mib#1{\mbox{\boldmath $#1$}}
\def\sla#1{\mbox{$#1\!\!\scriptstyle{/}$}}

\def\slak{\mbox{$k\!\!\!\scriptstyle{/}$}\,}
\def\bra#1{\langle #1 |} \def\ket#1{|#1\rangle}
\def\vev#1{\langle #1\rangle}

\nc{\barx}{\bar{x}}\nc{\pbarn}{\;\hbox {pb}}\nc{\fbarn}{\;\hbox {fb}}
\nc{\hc}{\hbox {h.c.}} \nc{\re}{\hbox {Re}} 
\nc{\mev}{\hbox {MeV}} \nc{\gev}{\;\hbox {GeV}}
\def\gesim{\lower0.5ex\hbox{$\:\buildrel >\over\sim\:$}}
\def\lesim{\lower0.5ex\hbox{$\:\buildrel <\over\sim\:$}}
\nc{\prd}[3]{{\it Phys.\ Rev.}\ {{\bf D{#1}} (#2), #3}}
\nc{\prl}[3]{{\it Phys.\ Rev.\ Lett.}\ {{\bf {#1}} (#2), #3}}
\nc{\plb}[3]{{\it Phys.\ Lett.}\ {{\bf B{#1}} (#2), #3}}
\nc{\npb}[3]{{\it Nucl.\ Phys.}\ {{\bf B{#1}} (#2), #3}}
\nc{\ptp}[3]{{\it Prog.\ Theor.\ Phys.}\ {{\bf {#1}} (#2), #3}}
\nc{\zfp}[3]{{\it Z.\ Phys.}\ {{\bf C{#1}} (#2), #3}}
\nc{\epj}[3]{{\it Eur.\ Phys.\ J.}\ {{\bf C{#1}} (#2), #3}}
\nc{\mpla}[3]{{\it Mod.\ Phys.\ Lett.}\ {{\bf A{#1}} (#2), #3}}
\nc{\rmp}[3]{{\it Rev.\ Mod.\ Phys.}\ {{\bf {#1}} (#2), #3}}
\nc{\ijmpa}[3]{{\it Int.\ J.\ Mod.\ Phys.}\
               {{\bf A{#1}} (#2), #3}}
\nc{\ttbar}{t\bar{t}}         \nc{\bbbar}{b\bar{b}}
\nc{\tanb}{\tan \beta}        \nc{\twbdec}{t\to W^+ b}
\nc{\tbwbdec}{\bar{t}\to W^- \bar{b}}
\nc{\epem}{e^+e^-}            \nc{\eett}{\epem \to \ttbar}
\nc{\sigeett}{\sigma_{e\bar{e}\to\ttbar}}
\nc{\wpwm}{W^+W^-}            \nc{\tbar}{\bar{t}}
\nc{\bbar}{\bar{b}}           \nc{\wpp}{W^+}
\nc{\mt}{m_t}    \nc{\mts}{m_t^2}   \nc{\mw}{m_W}    \nc{\mws}{m_W^2}
\nc{\mz}{m_Z}    \nc{\mzs}{m_Z^2}
\nc{\ttbardec}{\ttbar \to W^+W^-\bbbar}
\nc{\wwbb}{W^+W^-\bbbar}      \nc{\sm}{SM}
\nc{\cw}{\cos\theta_W}        \nc{\sw}{\sin\theta_W}
\nc{\sws}{\sin^2\theta_W}     \nc{\sig}{\sigma_{tot}}
\nc{\lp}{{\ell}^+}              \nc{\lm}{{\ell}^-}
\nc{\epsl}{\epsilon_L}        \nc{\cp}{C\!P}
\nc{\splus}{s_+}       \nc{\smin}{s_-}        \nc{\eps}{\epsilon}
\nc{\psp}{Ps_+}        \nc{\psm}{Ps_-}        \nc{\lsp}{ls_+}
\nc{\lsm}{ls_-}        \nc{\sss}{s_+s_-}      \nc{\m}{m_t}
\nc{\mq}{m_t^2}        \nc{\mr}{\frac{1}{\m}} \nc{\av}{A_{\gamma}}
\nc{\bv}{B_{\gamma}}   \nc{\az}{A_Z}          \nc{\bz}{B_Z}
\nc{\avs}{A_{\gamma}^2}\nc{\azs}{A_Z^2}       \nc{\bzs}{B_Z^2}
\nc{\dav}{\delta \! A_{\gamma}}   \nc{\dbv}{\delta \! B_{\gamma}}
\nc{\dcv}{\delta C_{\gamma}}      \nc{\ddv}{\delta \! D_{\gamma}}
\nc{\daz}{\delta \! A_Z}          \nc{\dbz}{\delta \! B_Z}
\nc{\dcz}{\delta C_Z}             \nc{\ddz}{\delta \! D_Z}
\nc{\dev}{\delta \! E_{\gamma}}   \nc{\dez}{\delta \! E_Z}
\nc{\dfv}{\delta \! F_{\gamma}}   \nc{\dfz}{\delta \! F_Z}
\nc{\rdav}{{\rm Re}(\delta \! A_{\gamma}) \:}
\nc{\rdbv}{{\rm Re}(\delta \! B_{\gamma}) \:}
\nc{\rdcv}{{\rm Re}(\delta C_{\gamma}) \:}
\nc{\rddv}{{\rm Re}(\delta \! D_{\gamma}) \:}
\nc{\rdaz}{{\rm Re}(\delta \! A_Z) \:}
\nc{\rdbz}{{\rm Re}(\delta \! B_Z) \:}
\nc{\rdcz}{{\rm Re}(\delta C_Z) \:}
\nc{\rddz}{{\rm Re}(\delta \! D_Z) \:}
\nc{\idav}{{\rm Im}(\delta \! A_{\gamma}) \:}
\nc{\idbv}{{\rm Im}(\delta \! B_{\gamma}) \:}
\nc{\idcv}{{\rm Im}(\delta C_{\gamma}) \:}
\nc{\iddv}{{\rm Im}(\delta \! D_{\gamma}) \:}
\nc{\idaz}{{\rm Im}(\delta \! A_Z) \:}
\nc{\idbz}{{\rm Im}(\delta \! B_Z) \:}
\nc{\idcz}{{\rm Im}(\delta C_Z) \:}
\nc{\iddz}{{\rm Im}(\delta \! D_Z) \:}
\nc{\cz}{(1+v_e^2)d\:\!'^2}         \nc{\ci}{v_ed\:\!'}
\nc{\ccz}{v_ed\:\!'^2}              \nc{\cci}{d\:\!'}
\nc{\lspace}{\;\;\;\;\;\;\;\;\;\;}  \nc{\llspace}{\lspace \lspace}
\nc{\beq}{\begin{equation}}   \nc{\eeq}{\end{equation}}
\nc{\bea}{\begin{eqnarray}}   \nc{\eea}{\end{eqnarray}}
\nc{\baa}{\begin{array}}      \nc{\eaa}{\end{array}}
\nc{\bit}{\begin{itemize}}    \nc{\eit}{\end{itemize}}
\nc{\ben}{\begin{enumerate}}  \nc{\een}{\end{enumerate}}
\nc{\bce}{\begin{center}}     \nc{\ece}{\end{center}}
\nc{\ocal}{{\cal O}}
\begin{document}
\pagestyle{empty} \setlength{\footskip}{2.0cm}
\setlength{\oddsidemargin}{0.5cm} \setlength{\evensidemargin}{0.5cm}
\renewcommand{\thepage}{-- \arabic{page} --}
\def\mib#1{\mbox{\boldmath $#1$}}
\def\bra#1{\langle #1 |}      \def\ket#1{|#1\rangle}
\def\vev#1{\langle #1\rangle} \def\dps{\displaystyle}
\nc{\tb}{\stackrel{{\scriptscriptstyle (-)}}{t}}
\nc{\bb}{\stackrel{{\scriptscriptstyle (-)}}{b}}
\nc{\fb}{\stackrel{{\scriptscriptstyle (-)}}{f}}
\nc{\pp}{\gamma \gamma}
\nc{\pptt}{\pp \to \ttbar}
   \def\thebibliography#1{\centerline{REFERENCES}
     \list{[\arabic{enumi}]}{\settowidth\labelwidth{[#1]}\leftmargin
     \labelwidth\advance\leftmargin\labelsep\usecounter{enumi}}
     \def\newblock{\hskip .11em plus .33em minus -.07em}\sloppy
     \clubpenalty4000\widowpenalty4000\sfcode`\.=1000\relax}\let
     \endthebibliography=\endlist
   \def\sec#1{\addtocounter{section}{1}\section*{\hspace*{-0.72cm}
     \normalsize\bf\arabic{section}.$\;$#1}\vspace*{-0.3cm}}
\vspace*{-1.7cm}
\noindent
%
\begin{flushright}
$\vcenter{
\hbox{{\footnotesize IFT-25-03\ \ \ FUT-03-02}}
\hbox{{\footnotesize UCRHEP-T364}}
\hbox{{\footnotesize TOKUSHIMA Report}}
\hbox{(hep-ph/0310159)}
}$
\end{flushright}

\vskip 0.6cm
\begin{center}
{\large\bf Probing Anomalous Top-Quark Couplings Induced}

\vskip 0.15cm
{\large\bf by Dim.6 Operators at Photon Colliders}
\end{center}

\vspace{0.2cm}
\begin{center}
\renewcommand{\thefootnote}{\alph{footnote})}
{\sc Bohdan GRZADKOWSKI$^{\:1),\:}$}\footnote{E-mail address:
\tt bohdan.grzadkowski@fuw.edu.pl},\ \
{\sc Zenr\=o HIOKI$^{\:2),\:}$}\footnote{E-mail address:
\tt hioki@ias.tokushima-u.ac.jp},

\vskip 0.15cm
{\sc Kazumasa OHKUMA$^{\:3),\:}$}\footnote{E-mail address:
\tt ohkuma@ccmails.fukui-ut.ac.jp}\ and\
{\sc Jos\'e WUDKA$^{\:4),\:}$}\footnote{E-mail address:
\tt jose.wudka@ucr.edu}
\end{center}

\vspace*{0.2cm}
\centerline{\sl $1)$ Institute of Theoretical Physics,\ Warsaw
University}
\centerline{\sl Ho\.za 69, PL-00-681 Warsaw, POLAND}
\centerline{\sl and}
\centerline{\sl CERN, Department of Physics}
\centerline{\sl Theory Division}
\centerline{\sl 1211 Geneva 23, Switzerland}

\vskip 0.2cm
\centerline{\sl $2)$ Institute of Theoretical Physics,\
University of Tokushima}
\centerline{\sl Tokushima 770-8502, JAPAN}

\vskip 0.2cm
\centerline{\sl $3)$ Department of Management Science,\
Fukui University of Technology}
\centerline{\sl Fukui 910-8505, JAPAN}

\vskip 0.2cm
\centerline{\sl $4)$ Department of Physics,\
University of California}
\centerline{\sl Riverside CA 92521-0413, USA}

\vspace*{0.6cm}
\centerline{ABSTRACT}

\vspace*{0.3cm}
\baselineskip=20pt plus 0.1pt minus 0.1pt
Possible anomalous top-quark couplings induced by
$SU(2)\times U(1)$ gauge-in\-var\-i\-ant dimension-6 effective
operators were studied in the process of $t\tbar$ productions
and decays at polarized $\gamma\gamma$ colliders. Two
$C\!P$-violating asymmetries, a linear-polarization asymmetry
and a circular-polarization asymmetry, were computed 
including both non-standard $t\bar{t}\gamma$ and
$\gamma\gamma H$ couplings. An optimal-observable analysis 
for the process $\gamma\gamma \to \ttbar \to {\ell}^\pm \cdots$
was performed in order to estimate the precision for
determination of all relevant non-standard couplings, including
the anomalous $tbW$ coupling.
\vspace*{0.4cm} \vfill



\newpage
\renewcommand{\thefootnote}{$\sharp$\arabic{footnote}}
\pagestyle{plain} \setcounter{footnote}{0}
\baselineskip=21.0pt plus 0.2pt minus 0.1pt

\sec{Introduction}

A lot of data have been accumulated on the top-quark ever since its
discovery \cite{top}. However, it still remains an open question whether
the top-quark couplings obey the Standard-Model (SM) scheme of the
electroweak forces or there exists a contribution from some
non-standard physics. Next-generation $\epem$ linear colliders are
expected to work as top-quark factories, and therefore a lot of
attention has been paid to study top-quark interactions through
$e\bar{e}\to t \bar{t}$ (see, e.g., \cite{Atwood:2001tu,Abe:2001gc}
and their reference lists).

An interesting option for $\epem$ machines could be that of photon-photon
collisions, where initial energetic photons are produced through
electron and laser-light backward scattering
\cite{Ginzburg:1981vm,Borden:1992qd}; such a collider presents remarkable
advantages for the study of $C\!P$ violation. In the case of 
$e\bar{e}$ collisions, the only initial states that are relevant
are $C\!P$-even states $\ket{e_{L/R}\bar{e}_{R/L}}$ under the usual
assumption that the electron mass can be neglected and that the
leading contributions to $\ttbar$ production come from $s$-channel
vector-boson exchanges. Therefore all
$C\!P$-violating observables must be constructed from
final-particle momenta/polarizations. In contrast, a
$\gamma\gamma$ collider offers a unique possibility of preparing
the polarization of the incident photon beams, which can be used
to construct $C\!P$-violating asymmetries without relying on final-state
information. 

This is why a number of authors have considered top-quark production
and decays in $\gamma\gamma$ collisions in order to study
{\it i}) Higgs-boson couplings to the top-quark and photon
\cite{Grzadkowski:1992sa}-\cite{Asakawa:2003dh}, or
{\it ii}) anomalous top-quark couplings to photon 
\cite{Choi:1995kp}-\cite{Poulose:1997xk}.
However, what is supposed to be observed in real experiments are
combined signals that originate both from the process of top-quark 
production, {\it and in addition}, from its decays.
In this sense, we have to conclude that in general
those previous papers are not realistic enough.
Therefore here we will consider $\gamma\gamma \to \ttbar \to
{\ell}^{\pm}X$ including all possible non-standard interactions
together (production and decay), and perform comprehensive analysis
as model-independently as possible within the effective-Lagrangian
framework of Buchm\"uller and Wyler \cite{Buchmuller:1986jz}.

The paper is organized as follows. In sec.2 we will briefly
describe a framework for our effective-Lagrangian approach.
Section 3 will be devoted to determination of the operator set
relevant for the $\ttbar$ production/decay processes, and the
Feynman rules they induce. Based on them we compute two
polarization asymmetries in sec.4, and carry out an
optimal-observable analysis aiming to determine all the unknown
parameters simultaneously in sec.5. Then we summarize our results
in the final section. For completeness, in the appendix we
collect formulas that are needed to describe the Stokes parameters
for the initial photon beams. There we also present some other
formulas which are necessary for the calculation of the cross
sections $\sigma(\gamma\gamma\to t\bar{t} \to {\ell}^{\pm}X)$.
Throughout this work, we use FORM \cite{FORM} for main algebraic
calculations.

\sec{Framework}

In this article we use a model-independent technique
based on effective low-energy Lagrangian
\cite{Buchmuller:1986jz,Arzt:gp} to describe
possible new-physics effects. In this approach, we are supposed
to consider the SM Lagrangian modified by the addition of a
series of $SU(2)\times U(1)$ gauge-invariant operators
whose coefficients parameterize the low-energy effects of the
underlying high-scale physics.

Assuming that the heavy degrees of freedom decouple implies that
the effective operators have coefficients suppressed by appropriate
inverse powers of ${\mit\Lambda}$, where ${\mit\Lambda}$ expresses
the energy scale of new physics \cite{decoupling}. If ${\mit\Lambda}
\gg v\sim 250\gev$
then the leading effects are generated by operators of mass-dimension
6 (the dimension 5 operator violates the lepton-number conservation
and therefore it is irrelevant hereafter) \cite{Buchmuller:1986jz}:
\begin{equation}
{\cal L}_{\rm eff}={\cal L}_{\rm SM}
+\frac1{{\mit\Lambda}^2}\sum_i (\alpha_i {\cal O}_i + {\rm h.c.}).
\end{equation}
If the high-scale theory is a weakly-coupled gauge theory, one can show
that coefficients $\alpha_i$ of the operators that may be generated at
the tree level of the underlying theory could be ${\cal O}(1)$ while
those that can appear only at the one-loop level of perturbative
expansion must be suppressed by at least the loop-factor $1/(4\pi)^2$
\cite{Arzt:gp}. 
Therefore we will assume all the couplings except those from the
SM tree level are small and take into account neither contributions
of order $1/{\mit\Lambda}^n$ with $n > 2$ nor the SM higher-order
terms.
%
%
Below we will refer
to application of ${\cal L}_{\rm eff}$ as to ``B\&W'' scenario
\cite{Buchmuller:1986jz}. Given our emphasis on top-quark physics the
effects of the first two fermion generations will be ignored.

In our framework, certain types of anomalous interactions are not included.
For instance, $\gamma\gamma Z$ couplings, which has been studied in
\cite{Poulose:1998sd}, is one of those couplings since it is not on the
list of $SU(3)\times SU(2)\times U(1)$ invariant dim.6 operators
(such an operator will appear with a suppression
factor of $ 1/{\mit\Lambda}^4$ and can be ignored). An
extended $t\bar{t}H$ coupling is not taken into account either, since
the other end of the Higgs propagator in $\gamma\gamma\to H\to t\bar{t}$
is a pure non-standard $\gamma\gamma H$ vertex, which requires that the
$t\bar{t}H$ coupling should come from the SM tree level within our
approximation.

Before going to the next section, there is an important comment in
order. Even though our approach is model independent, one should keep
in mind that we assume
${\mit\Lambda} \gg v$. Consequently for the process considered here our
results should not be directly applicable in the context of, e.g., two
Higgs-doublet model with extra scalar bosons having their masses of
${\cal O}(v)$. That kind of non-standard interactions would lead to
non-local form-factors that cannot be accommodated within our present
framework. This is because of the virtual and therefore off-shell
top-quark line that is present in the amplitudes for $\gamma \gamma
\to \ttbar$. Note that this is quite in contrast to $e\bar{e}\to\gamma/Z
\to t\bar{t}$ case, where for a given $\sqrt{s}$ we are able to write
down the most general invariant amplitude with constant form factors
because all the kinematic variables are fixed. 

\sec{Anomalous couplings from dim.6 operators}

Within the B\&W scenario, the following dim.6 operators
could contribute to the continuum top-quark production process $\pptt$:
\begin{equation}
\begin{array}{ll}
\ocal_{uB}=i\bar{u}\gamma_\mu D_\nu u B^{\mu\nu}, &
\ocal_{qB}=i\bar{q}\gamma_\mu D_\nu q B^{\mu\nu}, \\
\ocal_{qW}=i\bar{q}\tau^i\gamma_\mu D_\nu q W^{i\;\mu\nu}, &
\ocal_{uB}'=(\bar{q}\sigma^{\mu\nu}u)\tilde{\varphi} B_{\mu\nu}, \\
\ocal_{uW}
=(\bar{q}\sigma^{\mu\nu}\tau^i u)\tilde{\varphi} W_{i\;\mu\nu}, &
\end{array}
\label{prod-cont}
\end{equation}
where we have adopted the notation from ref.\cite{Buchmuller:1986jz}.
Each of the above operators contains both $C\!P$-violating and
$C\!P$-conserving parts.

These operators, however, are not independent. Using the
Bianchi identities and SM equations of motion we find:
\bea
&&\ocal_{uB}=-\frac{1}{4}[\:i{\mit\Gamma}_u ({\cal O}_{uB}^\prime
+ g^\prime {\cal O}_{u\varphi})
+ {\rm h.c.}\:]-i\frac{g^\prime}{2}{\cal O}_{\varphi u} \non\\
&&\phantom{\ocal_{uB}}
+ 4\mbox{\rm -fermion~operators}
+ {\rm total}\;\; {\rm derivative}, \\
&&\ocal_{qB}=\frac{1}{4}[\:i{\mit\Gamma}_u ({\cal O}_{uB}^\prime
+ g^\prime {\cal O}_{u\varphi})
+ i{\mit\Gamma}_d ({\cal O}_{dB}^\prime
+ g^\prime {\cal O}_{d\varphi})+ {\rm h.c.}\:] \non\\
&&\phantom{\ocal_{uB}}
-i\frac{g^\prime}{2}{\cal O}_{\varphi q}^{(1)}
+ 4\mbox{\rm -fermion~operators} + {\rm total}\;\;
{\rm derivative}, \\
&&\ocal_{qW}=\frac{1}{4}[\:i{\mit\Gamma}_u(\ocal_{uW}-g\ocal_{u\varphi})
+ i{\mit\Gamma}_d(\ocal_{dW}+g\ocal_{d\varphi})+ {\rm h.c.} \:]\non\\
&&\phantom{\ocal_{uB}}
-i\frac{g}{2}{\cal O}_{\varphi q}^{(3)}+ 4\mbox{\rm -fermion~operators}
+ {\rm total}\;\; {\rm derivative},
\label{red}
\eea
where ${\mit\Gamma}_{u,d}$ are the Yukawa couplings for up- and down-type
quarks, respectively.\footnote{
    Notice an omission of the term $i\varphi^\dagger\!
    \stackrel{\leftrightarrow}{D}_\beta\!\varphi/2$ in eq.(2.14) of
    \cite{Buchmuller:1986jz}.}
These relations imply that the operators $\ocal_{uB}$,
$\ocal_{qB}$ and $\ocal_{qW}$ are redundant and can
be dropped, which means the set of relevant operators is reduced to
\begin{equation}
\begin{array}{ll}
\ocal_{uB}'=(\bar{q}\sigma^{\mu\nu}u)\tilde{\varphi} B_{\mu\nu},
& \ocal_{uW}
=(\bar{q}\sigma^{\mu\nu}\tau^i u)\tilde{\varphi} W_{i\;\mu\nu}.
\end{array}
\label{prod-cont-eff}
\end{equation}
This reduction is important, as it allows to determine the minimal set of
operators that are relevant for the process considered here. In practice,
it can drastically decrease the effort that otherwise would be
necessary to obtain the final and correct result. In particular, it shows
that the contact interactions of the type $t\bar{t}\gamma\gamma$ should
be dropped.

On the other hand, the following operators contribute to $\pptt$
through the resonant $s$-channel Higgs-boson exchange:
\begin{equation}
\begin{array}{ll}
\ocal_{\varphi\tilde{W}}
=(\varphi^\dagger \varphi)\tilde{W}_{\mu\nu}^i W^{i\; \mu\nu}, &
\ocal_{\varphi W}
=(\varphi^\dagger \varphi)W_{\mu\nu}^i W^{i\; \mu\nu}/2, \\
\ocal_{\varphi\tilde{B}}
=(\varphi^\dagger \varphi)\tilde{B}_{\mu\nu} B^{\mu\nu}, &
\ocal_{\varphi B}
=(\varphi^\dagger \varphi)B_{\mu\nu} B^{\mu\nu}/2, \\
\ocal_{\tilde{W}B}
=(\varphi^\dagger\tau^i \varphi)\tilde{W}_{\mu\nu}^i B^{\mu\nu}, &
\ocal_{WB}
=(\varphi^\dagger\tau^i \varphi)W_{\mu\nu}^i B^{\mu\nu}.
\end{array}
\label{prod-res}
\end{equation}
The operators that contain the dual tensors
(e.g., $\tilde{B}_{\mu\nu}
\equiv \epsilon_{\mu\nu\alpha\beta}B^{\alpha\beta}/2$ with
$\epsilon_{0123}=+1$) are $C\!P$ odd while the remaining are
$C\!P$ even.

These operators lead to the following Feynman rules for on-shell
photons, which are necessary for our later calculations: \\ \\
(1) $C\!P$-conserving $t\bar{t}\gamma$ vertex
\begin{equation}
\frac{\sqrt{2}}{{\mit\Lambda}^2}v \alpha_{\gamma 1}\,
\slak\gamma_\mu,
\end{equation}
(2) $C\!P$-violating $t\bar{t}\gamma$ vertex
\begin{equation}
i\frac{\sqrt{2}}{{\mit\Lambda}^2}v \alpha_{\gamma 2}\,
\slak\gamma_\mu \gamma_5, \label{cpv-vertex}
\end{equation}
(3) $C\!P$-conserving $\gamma\gamma H$ vertex
\begin{eqnarray}
&&-\frac{4}{{\mit\Lambda}^2}v \alpha_{h1}\,
\bigl[\:
(k_1 k_2)g_{\mu\nu}-k_{1\nu}k_{2\mu}
\:\bigr],
\end{eqnarray}
(4) $C\!P$-violating $\gamma\gamma H$ vertex
\begin{eqnarray}
&&\frac{8}{{\mit\Lambda}^2}v \alpha_{h2}\,
k_1^\rho k_2^\sigma \epsilon_{\rho\sigma\mu\nu},
\end{eqnarray}
where $k$ and $k_{1,2}$ are incoming photon momenta, and
$\alpha_{\gamma 1,\gamma 2,h1,h2}$ are defined as
\begin{eqnarray}
&&\alpha_{\gamma 1}\equiv
\sin\theta_W{\rm Re}(\alpha_{uW})
+\cos\theta_W{\rm Re}(\alpha'_{uB}),
\\
&&\alpha_{\gamma 2}\equiv
\sin\theta_W{\rm Im}(\alpha_{uW})
+\cos\theta_W{\rm Im}(\alpha'_{uB}),
\\
&&\alpha_{h 1}\equiv
\sin^2\theta_W{\rm Re}(\alpha_{\varphi W})
+\cos^2\theta_W{\rm Re}(\alpha_{\varphi B}) 
-2\sin\theta_W \cos\theta_W{\rm Re}(\alpha_{WB}),
\\
&&\alpha_{h 2}\equiv
\sin^2\theta_W{\rm Re}(\alpha_{\varphi \tilde{W}})
+\cos^2\theta_W{\rm Re}(\alpha_{\varphi \tilde{B}}) 
-\sin\theta_W \cos\theta_W{\rm Re}(\alpha_{\tilde{W}B}).
\end{eqnarray}
In our notation, the standard-model $f\bar{f}\gamma$ coupling
is given by
\[
eQ_f \gamma_\mu,
\]
where $e$ is the proton charge and $Q_f$ is $f$'s electric charge
in $e$ unit (e.g., $Q_u = 2/3$).

The top-quark decay vertex is also affected by some dim.6 operators.
For the on-mass-shell $W$ boson it will be sufficient to consider
just the following $tbW$ amplitude since other possible terms do
not interfere with the SM tree-level vertex when $m_b$ is neglected:
\begin{equation}
{\mit\Gamma}^{\mu}_{Wtb}=-{g\over\sqrt{2}}\:
\bar{u}(p_b)\biggl[\,\gamma^{\mu} f_1^L P_L
-{{i\sigma^{\mu\nu}k_{\nu}}\over M_W}
f_2^R P_R\,\biggr]u(p_t),\label{ffdef}\\
\end{equation}
where $P_{L,R}\equiv (1\mp \gamma_5)/2$, and $f_1^L$ and $f_2^R$
are given by
\begin{eqnarray}
&&f^L_1=1+\frac{v}{{\mit\Lambda}^2}
  \Bigl[\:
\frac{m_t}{2}\alpha_{Du}
  -2v\alpha_{\varphi q}^{(3)}\:\Bigr], \\
&&f^R_2=-\frac{v}{{\mit\Lambda}^2}M_W
  \Bigl[\:\frac{4}{g}\alpha_{uW}
  +\frac12 \alpha_{Du}
\:\Bigr],  \label{f2R}
\end{eqnarray}
with $\alpha_{Du}$ and $\alpha_{\varphi q}^{(3)}$
being the coefficients of the following operators:\footnote{Note that
   there is another potential source of contribution, which may come 
   from ${\cal O}_{\bar D u}=(D_{\mu}\bar{q})u D^{\mu}\tilde{\varphi}$.
   However, this operator could be eliminated using equations of motion;
   therefore, it is neglected hereafter. We thank Ilya Ginzburg for
   pointing this to us.}.

\beq
{\cal O}_{Du}=(\bar{q}D_{\mu}u)D^{\mu}\tilde{\varphi},\;\;\;\;
{\cal O}_{\varphi q}^{(3)}
=i(\varphi^{\dagger}D_{\mu}\tau^i\varphi)(\bar{q}\gamma^{\mu}\tau^i q).
\eeq
Finally, the $\nu\ell W$ vertex is assumed to receive negligible contributions
from physics beyond the SM.


\sec{Polarization asymmetries}

We are now ready to calculate the cross section of $\gamma\gamma\to
t\bar{t}(\to {\ell}^{\pm}X)$. The work is straightforward and we
carried it out via FORM \cite{FORM} as mentioned in sec.1.
The analytical results are
however too long to show in this paper. We therefore refrain from
showing them here and we will present only some general formulas in
appendix A1. As a direct application of those calculations, in this
section we study $C\!P$-violating asymmetries. $C\!P$-violation
phenomena would be an interesting indication of some new physics
since SM contribution is negligible in the top-quark couplings.

As mentioned in the Introduction, we can form $C\!P$-violating
asymmetries by adjusting initial-state polarizations. 
These are
characterized by the initial electron and positron
longitudinal-polarizations 
$P_e$ and $P_{\bar{e}}$, the average
helicities of the initial-laser-photons $P_{\gamma}$ and
$P_{\tilde{\gamma}}$, and their maximum average linear-polarizations
$P_t$ and $P_{\tilde{t}}$ with the azimuthal angles $\varphi$ and
$\tilde{\varphi}$
(defined the same way as in \cite{Ginzburg:1981vm}).
$P_{\gamma,t}$ and $P_{\tilde{\gamma},\tilde{t}}$ have to satisfy
\begin{equation}
0 \leq P_{\gamma}^2 + P_t^2 \leq 1,
\ \ \ \ \ \ \ \ \
0 \leq P_{\tilde{\gamma}}^2 + P_{\tilde{t}}^2 \leq 1.
\end{equation}

\begin{figure}[h] 
  \begin{center}
 \includegraphics[width=10cm,clip]{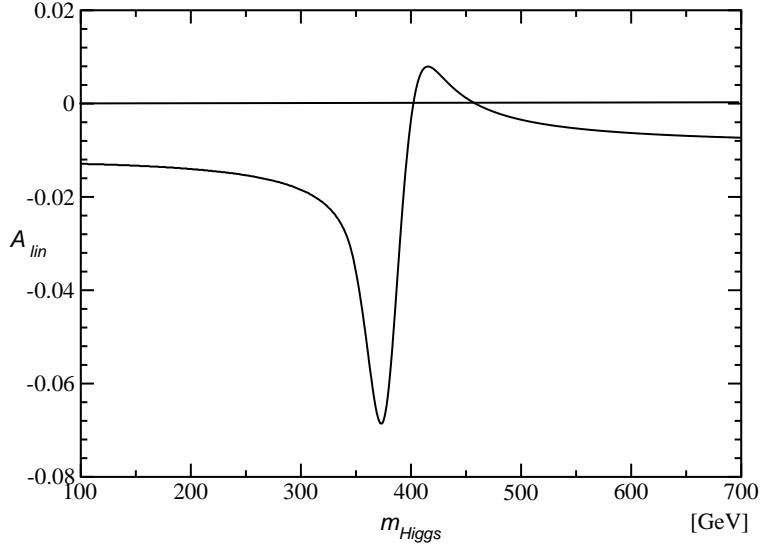}
  \end{center}
 \caption{$C\!P$-violating linear-polarization asymmetry
 $A_{lin}$ as a function of the Higgs-boson mass
 computed for $\sqrt{s_{e\bar{e}}}=500$ GeV, ${\mit\Lambda}=1$ TeV,
 $y_0=\tilde{y}_0=4.828$, $\alpha_{\gamma 2}=\alpha_{h 2}=0.1$,
 $P_e=P_{\bar{e}}=1$, $P_t=P_{\tilde{t}}=P_{\gamma}=P_{\tilde{\gamma}}=
 1/\sqrt{2}$ and $\chi\equiv
 \varphi-\tilde{\varphi}=\pm\pi/4$.}\label{Alin}
 %
 \end{figure}

\begin{figure}[t]  
  \begin{center}
 \includegraphics[width=10cm,clip]{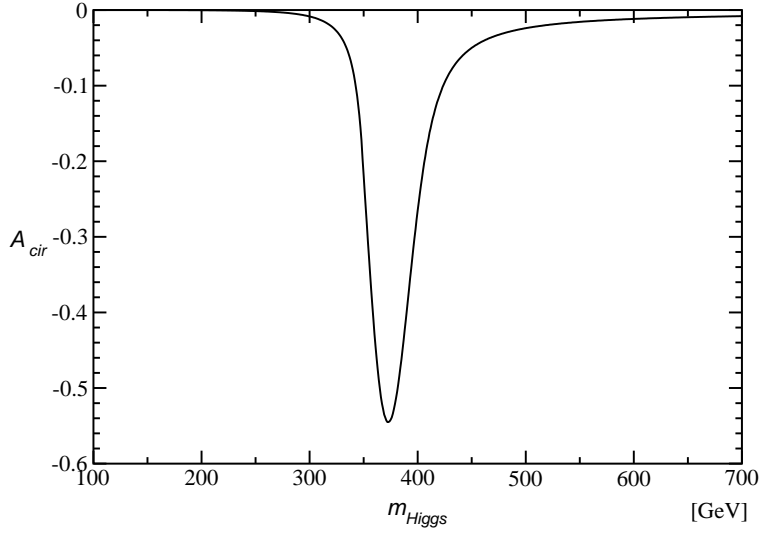}
  \end{center}
 \caption{$C\!P$-violating circular-polarization asymmetry
 $A_{cir}$ as a function of the Higgs-boson mass
 computed for $\sqrt{s_{e\bar{e}}}=500$ GeV, ${\mit\Lambda}=1$ TeV,
 $y_0=\tilde{y}_0=4.828$, $\alpha_{h 2}=0.1$, and $P_e=P_{\bar{e}}=
 P_{\gamma}=P_{\tilde{\gamma}}=\pm 1$.}\label{Acir}
 \end{figure}

Specifically we consider the following
$C\!P$-violating asymmetries:\footnote{These were used previously in
    \cite{Gounaris:1997ef} in order to study the $C\!P$ property of
    the Higgs boson produced in $\gamma\gamma \to H$. $A_{lin}$ was
    also used in \cite{Choi:1995kp,Baek:1997ib} for studying the
    $C\!P$-violating $t\bar{t}\gamma$ coupling.}
\begin{equation}
A_{lin}
\equiv \frac{\sigma(\chi=+\pi/4)-\sigma(\chi=-\pi/4)}
{\sigma(\chi=+\pi/4)+\sigma(\chi=-\pi/4)}
\end{equation}
and
\begin{equation}
A_{cir}
\equiv \frac{\sigma(++)-\sigma(--)}{\sigma(++)+\sigma(--)},
\end{equation}
where $\sigma(\chi=\pm\pi/4)$ means the total cross section of
$\gamma\gamma\to t\bar{t}$ with $P_e=P_{\bar{e}}=1$, $P_t=P_{\tilde{t}}=
P_{\gamma}=P_{\tilde{\gamma}}=1/\sqrt{2}$ and $\chi\equiv
\varphi-\tilde{\varphi}=\pm \pi/4$, while $\sigma(\pm\pm)$ corresponds
to the one with $P_e=P_{\bar{e}}=P_{\gamma}=P_{\tilde{\gamma}}=\pm 1$
(which means $P_t=P_{\tilde{t}}=0$). They were computed as functions of
the Higgs-boson mass for
\[
\sqrt{s_{e\bar{e}}}=500\ {\rm GeV},\ \ \
{\mit\Lambda}=1\ {\rm TeV},\ \ \ y_0=\tilde{y}_0=4.828,
\]
\[
\alpha_{\gamma 2}=\alpha_{h 2}=0.1
\]
and using the SM expression for the Higgs-boson width.
The results are shown in Figs.\ref{Alin} and \ref{Acir}.

\vfill

\newpage
\vspace*{0.3cm}

\vskip 0.3cm
The sharp peak in each asymmetry comes, of course, from the Higgs-boson
pole, where the signal will be easily observed. The resonance region has
been already studied in great details in existing literature
\cite{Grzadkowski:1992sa}-\cite{Asakawa:2003dh}. Here we
will study the possibility of extracting a
meaningful signal when the average
$\sqrt{s_{\gamma\gamma}}$ is far from the Higgs-boson mass. 
For the parameters adopted above one can
find approximate formulas that illustrate sensitivity to
$C\!P$-violating coefficients $\alpha_{\gamma 2}$ and $\alpha_{h 2}$:
\\
(1) $m_H=$ 100 GeV
\begin{eqnarray}
&&A_{lin}
=-(8.8\,\alpha_{\gamma 2} + 4.2\,\alpha_{h 2})\times 10^{-2}, \\
&&A_{cir}
=-5.2\times 10^{-6}\,\alpha_{h 2}.
\end{eqnarray}
(2) $m_H=$ 300 GeV
\begin{eqnarray}
&&A_{lin}
=-(8.8\,\alpha_{\gamma 2} + 11\,\alpha_{h 2})\times 10^{-2}, \\
&&A_{cir}
=-8.5\times 10^{-2}\,\alpha_{h 2}.
\end{eqnarray}
(3) $m_H=$ 500 GeV
\begin{eqnarray}
&&A_{lin}
=-(8.8\,\alpha_{\gamma 2} - 8.1\,\alpha_{h 2})\times 10^{-2}, \\
&&A_{cir}
=-0.24\,\alpha_{h 2}.
\end{eqnarray}
Note that because of the $\gamma_5$ factor  in the $C\!P$-violating
$\ttbar \gamma$ vertex (\ref{cpv-vertex}) its interference with
the SM amplitude vanishes when the initial photons have pure
circular polarizations \cite{Choi:1995kp}. This is why the
$\alpha_{\gamma 2}$ contribution to $A_{cir}$ is zero.

Since the asymmetries are all small, their expected statistical
errors in actual measurements can be computed by
\begin{equation}
{\mit\Delta}A_{lin,cir}\simeq 1/\sqrt{N_{t\bar{t}}}
\end{equation}
and consequently their statistical significances are estimated as
\begin{equation}
N_{SD}\equiv |A|/{\mit\Delta}A \simeq |A|\sqrt{N_{t\bar{t}}}.
\end{equation}

We estimate $N_{t\bar{t}}\sim$ 36,000 and 21,000 events for
the linear and circular polarizations respectively assuming that 
a luminosity of $L_{e\bar{e}}^{\rm eff}\equiv \epsilon L_{e\bar{e}}=500\ 
{\rm fb}^{-1}$ can be reached for each mode
($\epsilon$ denotes the relevant detection efficiency
and $L_{e\bar{e}}$ is the integrated luminosity). In this case
deviations from the SM will be observable provided the
coefficients $\alpha_{\gamma 2}$ and $\alpha_{h 2}$ are of the
order of $0.1$.

\sec{Optimal-observable analysis}

Let us briefly summarize the main points of this method first.
Suppose we have a differential cross section of the following form
\[
\frac{d\sigma}{d\phi}=f_0(\phi)+\sum_i c_i f_i(\phi),
\]
where $f_0(\phi)$ is the SM differential cross section, while
$f_i(\phi)$'s are known functions (of the same order of magnitude
as $f_0(\phi)$)
of the location in final-state phase-space $\phi$, and $c_i$
are model-dependent dimensionless coefficients expressing non-standard
contributions. We assume that $c_i$ are so small that we can safely
drop ${\cal O}(c_i^2)$ contributions. The goal would be to determine
$c_i$'s. It can be done by using appropriate weighting-functions
$w_i(\phi)$ such that
\[
c_i
= \Bigl[\int \! d\phi\,w_i(\phi)\frac{d\sigma}{d\phi}\:\Bigr]
\Big/\int \! d\phi\,f_0(\phi).
\]
(Note here that $d\sigma/d\phi$ is not always a properly normalized
probability distribution.) Generally, different choices for
$w_i(\phi)$ are possible, but there is a unique choice which
minimizes the resultant statistical error. Such weighting
functions are given by \cite{optimal}
\begin{equation}
w_i(\phi)
=I_0 \sum_j X_{ij}\bigl[\,f_j(\phi)/f_0(\phi)
-I_j/I_0\,\bigr]\,,  \label{wi}
\end{equation}
where $X_{ij}$ is the inverse matrix of $M_{ij}$ defined
as$\,$\footnote{If we made the corresponding matrix including
    the SM term, it would become ${\cal M}_{ij}=
    \int d\phi f_i(\phi)f_j(\phi)/f_0(\phi)$, where the zero-th
    component is the SM contribution. This matrix is what we have used
    in our previous papers. It is a simple linear-algebra exercise to
    show that this ${\cal M}_{ij}$ and $M_{ij}$ in eq.(\ref{M_def})
    both give the same $X_{ij}$ when restricted to the non-SM components.}
\begin{equation}
M_{ij}\equiv \int \! d\phi\,f_i(\phi)f_j(\phi)/f_0(\phi)
- I_i I_j/I_0\,,
\label{M_def}
\end{equation}
and
\[
I_0\equiv \int \!d\phi\, f_0(\phi)\,, \lspace
I_i\equiv \int \!d\phi\, f_i(\phi)\,.
\]
For the weighting functions chosen in the above manner, the
covariance matrix $V_{ij}$ that corresponds to $w_i(\phi)$ becomes
\beq
V_{ij} =I_0 X_{ij}/N + {\cal O}(c_i)\,,
\label{cov}
\eeq
where $N$ is the total number of collected events. Its diagonal
elements give expected statistical uncertainty for measurements
of $c_i$ as
\beq
|{\mit\Delta}c_i|=\sqrt{V_{ii}}.
\eeq

We are going to apply this technique to the final-lepton angular
and energy distribution of $\gamma\gamma\to t\bar{t}\to{\ell}^+X$
\begin{eqnarray}
&&\frac{d\sigma}{dE_{\ell} d\cos\theta_{\ell}}
=f_{\rm SM}(E_{\ell}, \cos\theta_{\ell})
 + \alpha_{\gamma 1} f_{\gamma 1}(E_{\ell}, \cos\theta_{\ell})
 + \alpha_{\gamma 2} f_{\gamma 2}(E_{\ell}, \cos\theta_{\ell}) \non\\
&&\phantom{\frac{d\sigma}{dE_{\ell}}}\ \ \
 + \alpha_{h1} f_{h1}(E_{\ell}, \cos\theta_{\ell})
 + \alpha_{h2} f_{h2}(E_{\ell}, \cos\theta_{\ell})
 + \alpha_d f_d(E_{\ell}, \cos\theta_{\ell})~~~
\label{distribution}
\end{eqnarray}
in the $e\bar{e}$-CM frame. Here $f_{\rm SM}$ denotes the standard-model
contribution, $f_{\gamma 1,\gamma 2}$ describe, respectively, the anomalous
$C\!P$-conserving- and
$C\!P$-violating-$t\bar{t}\gamma$-vertices contributions,
$f_{h1,h2}$  those generated by the anomalous $C\!P$-conserving and
$C\!P$-violating $\gamma\gamma H$-vertices,
and $f_d$ that by the anomalous $tbW$-vertex  with
\[
\alpha_d = {\rm Re}(f_2^R).
\]
See the Appendix for details of our calculation framework.

The covariance matrix $V_{ij}\propto X_{ij}$ determines our ability
to measure the coefficients $c_i$. However in the case considered 
here, it turns out that our results for $V_{ij}$ are very unstable:
even a tiny fluctuation of $M_{ij}$ changes $X_{ij}$
significantly. This indicates that some of $f_i$ have similar
shapes$\,$\footnote{Note that if two $f_i$ functions were proportional to
    each other then the matrix $M_{ij}$ would have a vanishing
    determinant, and therefore its inverse $X_{ij}$ could not
    be determined.}\
and therefore their coefficients cannot be easily 
disentangled. The only option in such a case is to refrain from  
determining all the couplings at once through
this process alone. Therefore hereafter we assume that some of
$c_i$'s have been measured in other processes (e.g. in
$e\bar{e}\to t\bar{t}\to{\ell}^{\pm}X$).

In order to carry out these studies numerically, we first give full
elements of
\[
{\cal M}_{IJ}=\int dE_{\ell}d\cos\theta_{\ell}\,
f_I(E_{\ell}, \cos\theta_{\ell})f_J(E_{\ell}, \cos\theta_{\ell})/
f_1(E_{\ell}, \cos\theta_{\ell}),
\]
where $I,J=1,\cdots, 6$ correspond to SM, $\gamma 1$, $\gamma 2$, 
$h1$, $h2$ and $d$ respectively as our basis. The $(i,j)$ elements
$(i,j=2\sim 6)$ of ${\cal M}$'s inverse matrix coincide with $X_{ij}$
appearing in eq.(\ref{wi}) as mentioned in footnote 3. 

\vskip 0.5cm \noindent
(1) Linear Polarization

We took $P_e=P_{\bar{e}}=1$,
$P_t =P_{\tilde{t}}=P_{\gamma}=P_{\tilde{\gamma}}=1/\sqrt{2}$
and $\chi=\pi/4$, which were used to compute $A_{lin}$ in the
previous section as typical linear-polarization parameters.

\noindent
1-1) $m_H=100$ GeV
\begin{equation}
 \begin{array}{lll}
  {\cal M}_{11}= 0.810\cdot 10^{1},&  {\cal M}_{12}= 0.173\cdot 10^{2},&  
  {\cal M}_{13}=-0.703, \\
  {\cal M}_{14}=-0.319\cdot 10^{1},&  {\cal M}_{15}=-0.337,&
  {\cal M}_{16}= 0, \\
  {\cal M}_{22}= 0.377\cdot 10^{2},&  {\cal M}_{23}=-0.153\cdot 10^{1},&
  {\cal M}_{24}=-0.627\cdot 10^{1}, \\
  {\cal M}_{25}=-0.730,&              {\cal M}_{26}= 0.125\cdot 10^{1},&
  {\cal M}_{33}= 0.638\cdot 10^{-1}, \\
  {\cal M}_{34}= 0.248,            &  {\cal M}_{35}= 0.305\cdot 10^{-1},&
  {\cal M}_{36}=-0.683\cdot 10^{-1}, \\
  {\cal M}_{44}= 0.183\cdot 10^{1},&  {\cal M}_{45}= 0.122,&
  {\cal M}_{46}= 0.135\cdot 10^{1}, \\
  {\cal M}_{55}= 0.149\cdot 10^{-1},& {\cal M}_{56}=-0.263\cdot 10^{-1},&
  {\cal M}_{66}= 0.392\cdot 10^{1}.
 \end{array}
\end{equation}

\noindent
1-2) $m_H=300$ GeV
\begin{equation}
 \begin{array}{lll}
  {\cal M}_{11}= 0.810\cdot 10^{1},&  {\cal M}_{12}= 0.173\cdot 10^{2},&
  {\cal M}_{13}=-0.703, \\
  {\cal M}_{14}=-0.790\cdot 10^{1},&  {\cal M}_{15}=-0.152\cdot 10^{1},&
  {\cal M}_{16}= 0, \\
  {\cal M}_{22}= 0.377\cdot 10^{2},&  {\cal M}_{23}=-0.153\cdot 10^{1},&
  {\cal M}_{24}=-0.156\cdot 10^{2}, \\
  {\cal M}_{25}=-0.314\cdot 10^{1},&  {\cal M}_{26}= 0.125\cdot 10^{1},&
  {\cal M}_{33}= 0.638\cdot 10^{-1}, \\
  {\cal M}_{34}= 0.616,            &  {\cal M}_{35}= 0.128,&
  {\cal M}_{36}=-0.683\cdot 10^{-1}, \\
  {\cal M}_{44}= 0.112\cdot 10^{2},&  {\cal M}_{45}= 0.176\cdot 10^{1},&
  {\cal M}_{46}= 0.332\cdot 10^{1}, \\
  {\cal M}_{55}= 0.309,            &  {\cal M}_{56}= 0.260,&
  {\cal M}_{66}= 0.392\cdot 10^{1}.
 \end{array}
\end{equation}

\noindent
1-3) $m_H=500$ GeV
\begin{equation}
 \begin{array}{lll}
  {\cal M}_{11}= 0.810\cdot 10^{1},&  {\cal M}_{12}= 0.173\cdot 10^{2},&
  {\cal M}_{13}=-0.703, \\
  {\cal M}_{14}= 0.374\cdot 10^{1},&  {\cal M}_{15}=-0.221\cdot 10^{1},&
  {\cal M}_{16}= 0, \\
  {\cal M}_{22}= 0.377\cdot 10^{2},&  {\cal M}_{23}=-0.153\cdot 10^{1},&
  {\cal M}_{24}= 0.737\cdot 10^{1}, \\
  {\cal M}_{25}=-0.425\cdot 10^{1},&  {\cal M}_{26}= 0.125\cdot 10^{1},&
  {\cal M}_{33}= 0.638\cdot 10^{-1}, \\
  {\cal M}_{34}=-0.293,            &  {\cal M}_{35}= 0.166,&
  {\cal M}_{36}=-0.683\cdot 10^{-1}, \\
  {\cal M}_{44}= 0.251\cdot 10^{1},&  {\cal M}_{45}=-0.165\cdot 10^{1},&
  {\cal M}_{46}=-0.155\cdot 10^{1}, \\
  {\cal M}_{55}= 0.110\cdot 10^{1},&  {\cal M}_{56}= 0.122\cdot 10^{1},&
  {\cal M}_{66}= 0.392\cdot 10^{1}.
 \end{array}
\end{equation}

\noindent
(2) Circular Polarization

What we took as circular-polarization parameters are also those
used for $A_{cir}$: 
$P_e=P_{\bar{e}}=P_{\gamma}=P_{\tilde{\gamma}}=1$.

\noindent
2-1) $m_H=100$ GeV
\begin{equation}
 \begin{array}{lll}
 {\cal M}_{11}= 0.460\cdot 10^{1},& {\cal M}_{12}= 0.996\cdot 10^{1},&
 {\cal M}_{13}= 0, \\
 {\cal M}_{14}=-0.152\cdot 10^{1},& {\cal M}_{15}=-0.240\cdot 10^{-4},&
 {\cal M}_{16}= 0, \\
 {\cal M}_{22}= 0.219\cdot 10^{2},& {\cal M}_{23}= 0        ,&
 {\cal M}_{24}=-0.306\cdot 10^{1}, \\
 {\cal M}_{25}=-0.481\cdot 10^{-4},& {\cal M}_{26}= 0.611,&
 {\cal M}_{33}= 0, \\
 {\cal M}_{34}= 0,                & {\cal M}_{35}= 0        ,&
 {\cal M}_{36}= 0, \\
 {\cal M}_{44}= 0.748,            & {\cal M}_{45}= 0.122\cdot 10^{-4},&
 {\cal M}_{46}= 0.660, \\
 {\cal M}_{55}= 0.198\cdot 10^{-9},& {\cal M}_{56}= 0.114\cdot 10^{-4},&
 {\cal M}_{66}= 0.229\cdot 10^{1}.
 \end{array}
\end{equation}

\noindent
2-2) $m_H=300$ GeV
\begin{equation}
 \begin{array}{lll}
 {\cal M}_{11}= 0.460\cdot 10^{1},& {\cal M}_{12}= 0.996\cdot 10^{1},&
 {\cal M}_{13}= 0, \\
 {\cal M}_{14}=-0.392\cdot 10^{1},& {\cal M}_{15}=-0.389,&
 {\cal M}_{16}= 0, \\
 {\cal M}_{22}= 0.219\cdot 10^{2},& {\cal M}_{23}= 0        ,&
 {\cal M}_{24}=-0.793\cdot 10^{1}, \\
 {\cal M}_{25}=-0.781,            & {\cal M}_{26}= 0.611,&
 {\cal M}_{33}= 0, \\
 {\cal M}_{34}= 0,                & {\cal M}_{35}= 0        ,&
 {\cal M}_{36}= 0, \\
 {\cal M}_{44}= 0.497\cdot 10^{1},& {\cal M}_{45}= 0.502,&
 {\cal M}_{46}= 0.171\cdot 10^{1}, \\
 {\cal M}_{55}= 0.511\cdot 10^{-1},            & {\cal M}_{56}= 0.181,&
 {\cal M}_{66}= 0.229\cdot 10^{1}.
 \end{array}
\end{equation}

\noindent
2-3) $m_H=500$ GeV
\begin{equation}
 \begin{array}{lll}
 {\cal M}_{11}= 0.460\cdot 10^{1},& {\cal M}_{12}= 0.996\cdot 10^{1},&
 {\cal M}_{13}= 0, \\
 {\cal M}_{14}= 0.168\cdot 10^{1},& {\cal M}_{15}=-0.110\cdot 10^{1},&
 {\cal M}_{16}= 0, \\
 {\cal M}_{22}= 0.219\cdot 10^{2},& {\cal M}_{23}= 0        ,&
 {\cal M}_{24}= 0.338\cdot 10^{1}, \\
 {\cal M}_{25}=-0.221\cdot 10^{1},& {\cal M}_{26}= 0.611,&
 {\cal M}_{33}= 0, \\
 {\cal M}_{34}= 0,                & {\cal M}_{35}= 0        ,&
 {\cal M}_{36}= 0, \\
 {\cal M}_{44}= 0.920,            & {\cal M}_{45}=-0.634,&
 {\cal M}_{46}=-0.728, \\
 {\cal M}_{55}= 0.436,            & {\cal M}_{56}= 0.537,&
 {\cal M}_{66}= 0.229\cdot 10^{1}.
 \end{array}
\end{equation}
%
All the elements ${\cal M}_{ij}$ above are given in units of
${\rm fb}$. In these results, the third components of ${\cal M}$
for the circular polarization vanish as was mentioned in sec.4. Also,
in accordance with the decoupling theorem
shown in \cite{Grzadkowski:2002gt}, the elements ${\cal M}_{16}$ are
always zero within our approximation of 
neglecting contributions quadratic in non-standard interactions
and treating the decaying $t$ and $W$ as on-shell particles.

When estimating the statistical uncertainty in simultaneous
measurements of, e.g., $\alpha_{\gamma 1}$ and $\alpha_{h 1}$
(assuming all other coefficients are known), we  need only 
the components with indices 1, 2 and 4. Let us express the
resultant uncertainties as ${\mit\Delta}\alpha_{\gamma 1}^{[3]}$
and ${\mit\Delta}\alpha_{h 1}^{[3]}$, where ``3" means that we
took account of the input ${\cal M}_{ij}$ up to three decimal places.
In order to see how stable the results are, we also computed
${\mit\Delta}\alpha_{\gamma 1}^{[2]}$ and
${\mit\Delta}\alpha_{h 1}^{[2]}$ by rounding ${\cal M}_{ij}$ off to two
decimal places. Then, if both of the deviations
$|{\mit\Delta}\alpha_{\gamma 1,h 1}^{[3]}
-{\mit\Delta}\alpha_{\gamma 1,h 1}^{[2]}|/
{\mit\Delta}\alpha_{\gamma 1,h 1}^{[3]}$ are less than 10 \%, we
adopted the results as stable solutions. Therefore, the ambiguity
of the following results ${\mit\Delta}\alpha_{i}$, which are
${\mit\Delta}\alpha_{i}^{[3]}$, is at most 10 \%.

\vskip 0.2cm \noindent
1) Linear Polarization

What we obtained as stable solutions for linear polarization
are \\
$\bullet$ Independent of $m_H$
\begin{equation}
{\mit\Delta} \alpha_{\gamma 2}= 73/\sqrt{N_{\ell}},\ \ \ \
{\mit\Delta} \alpha_{d}= 1.9/\sqrt{N_{\ell}}, \label{Ri}
\end{equation}
$\bullet$ $m_H=$ 100 GeV
\begin{equation}
{\mit\Delta} \alpha_{h2}= 107/\sqrt{N_{\ell}},\ \ \ \
{\mit\Delta} \alpha_{d}= 1.6/\sqrt{N_{\ell}},
\end{equation}
$\bullet$  $m_H=$ 300 GeV
\begin{equation}
{\mit\Delta} \alpha_{h1}= 3.4/\sqrt{N_{\ell}},\ \ \ \
{\mit\Delta} \alpha_{d}= 3.2/\sqrt{N_{\ell}},
\end{equation}
where $N_{\ell}\simeq 4,000$ for $L^{\rm eff}_{e\bar{e}}
=500$ fb$^{-1}$. 

\vskip 0.3cm \noindent
2) Circular Polarization

Stable solutions which we found for circular polarization
are \\
$\bullet$ $m_H=$ 100 GeV
\begin{equation}
{\mit\Delta} \alpha_{h1}= 9.0/\sqrt{N_{\ell}},\ \ \ \
{\mit\Delta} \alpha_{d}= 3.0/\sqrt{N_{\ell}},
\end{equation}
$\bullet$ $m_H=$ 300 GeV
\begin{equation}
{\mit\Delta} \alpha_{h1}= 3.5/\sqrt{N_{\ell}},\ \ \ \
{\mit\Delta} \alpha_{d}= 3.0/\sqrt{N_{\ell}},
\end{equation}
\begin{equation}
{\mit\Delta} \alpha_{h2}= 35/\sqrt{N_{\ell}},\ \ \ \
{\mit\Delta} \alpha_{d}= 3.1/\sqrt{N_{\ell}},
\end{equation}
$\bullet$ $m_H=$ 500 GeV
\begin{equation}
{\mit\Delta} \alpha_{h1}= 7.7/\sqrt{N_{\ell}},\ \ \ \
{\mit\Delta} \alpha_{d}= 2.8/\sqrt{N_{\ell}},
\end{equation}
\begin{equation}
{\mit\Delta} \alpha_{h2}= 10/\sqrt{N_{\ell}},\ \ \ \
{\mit\Delta} \alpha_{d}= 2.8/\sqrt{N_{\ell}}, \label{Rf}
\end{equation}
where $N_{\ell}\simeq 2,300$ for $L^{\rm eff}_{e\bar{e}}
=500$ fb$^{-1}$.\footnote{We used the tree-level SM formula
   for computing $N_{\ell}$, so we have the same $N_{\ell}$ for
   different $m_H$.}

Unfortunately, we did not find any stable solution including
${\mit\Delta} \alpha_{\gamma 1}$. We therefore have to look
for other suitable processes to determine this parameter.
The precision of $\alpha_{\gamma 2}$ is not satisfactory
either, but since this is a $C\!P$-violating parameter, we
will be able to get some information on it from the $C\!P$
asymmetries which we gave in the previous section.

The above results are for ${\mit\Lambda}=$ 1 TeV. When one takes
the new-physics scale to be ${\mit\Lambda}'=\lambda{\mit\Lambda}$,
then all the above results (${\mit\Delta}\alpha_i$) are replaced
with ${\mit\Delta}\alpha_i/\lambda^2$, which means that the
right-hand sides of eqs.(\ref{Ri})--(\ref{Rf}) are multiplied by
$\lambda^2$.

It should be mentioned here that the above estimation of
expected errors does not take into account a possible
background. However the top-quark tagging in the one-lepton
and six-jets final state is relatively straightforward and
therefore should not increase the uncertainties found above very
dramatically. Of course, in the real experiment, one will have
to redo the analysis taking into account not only the background,
but also all the systematic errors which are unknown at present.
Therefore the fully realistic analysis cannot be performed
at this moment.

%
%
%

\sec{Summary and discussion}

We have studied here beyond-the-SM effects in the process $\pp
\to \ttbar (\to \ell^+ X)$ for arbitrarily-polarized photon
beams, taking advantage of the fact that one can control
polarizations of the incoming photon beams.
Non-standard interactions have been parameterized
through dim.6 local and gauge symmetric effective operators
\`a la Buchm\"uller and Wyler \cite{Buchmuller:1986jz}
toward the first realistic comprehensive model-independent
analysis of the process. We listed all the necessary operators
and the corresponding Feynman rules. It was shown that in 
the list of operators that seem to contribute there are some
which are redundant and should be dropped in the analysis.
Assuming
that those new-physics (NP) effects are small, we have kept only
terms linear in modification of the SM tree vertices.

We first computed two $C\!P$-violating polarization asymmetries
for linear and circular photon polarizations, and found there are
good chances of detecting their signals. We then applied the
optimal-observable technique to the final-lepton-momentum
distribution, and estimated statistical significances of measuring
each NP-parameters. Unfortunately, we had to conclude that it is
never realistic to try to determine all the independent
NP-parameters at once through $\gamma\gamma\to t\bar{t}\to
{\ell}^{\pm}X$ alone, but still we would be able to perform
useful analysis if we could utilize complementary information
collected in other independent processes.

We have not discussed here a possible background.
This is mainly because identifying the $t\bar{t}$ final state
in semileptonic-hadronic decays is easy, since we always have
a very energetic charged lepton and the other (anti)top quark
decaying purely hadronically with an invariant mass of $m_t$.
That is, the tagging in this particular final state will
be relatively easy.
However, some comments on background estimation will be helpful.
Although the most serious background is $W$-boson pair
productions and indeed its total cross section could be larger
than $\sigma_{tot}(t\bar{t})$, a dedicated simulation
study \cite{Takahashi:px} has shown
that $t\bar{t}$ events can be selected with signal-to-background
ratio of 10 by imposing appropriate invariant-mass constraints
on the final-particle momenta.

It should be stressed that for the purpose of future data analysis
it is unavoidable to have a tool which would allow consistently to
control all possible effects. For instance it is conceivable
that some non-standard effects from the top-quark decays would
mimic another non-standard interactions from the $t\bar{t}$
production process, therefore without an analysis that allows
to control all those contribution any meaningful data analysis
would be impossible.

Finally, it should be emphasized here that the effective-operator
strategy adopted in this article is valid only for ${\mit\Lambda}
\gg v\sim 250 \gev$, in contrast with 
$\epem \to t\bar{t}\to {\ell}^{\pm}X$. 
Should the reaction 
$\gamma \gamma \to \ttbar \to {\ell}^{\pm}X$
exhibit a deviation from the SM predictions  that cannot be
described properly within this framework, this would be an
indication of a low-energy beyond-the-SM physics, e.g., two
Higgs-doublet model with relatively low mass scale of new
scalar degrees of freedom.

\vspace{0.6cm}
\centerline{ACKNOWLEDGMENTS}

\vspace{0.3cm}
One of us (Z.H.) would like to thank Tohru Takahashi for very useful
discussion, Isamu Watanabe for showing some results of their numerical
computations, and Eri Asakawa and Saurabh Rindani for kind
correspondence on their works.
B.G. thanks Mikolaj Misiak for useful discussions concerning effective
Lagrangians.
This work is supported in part by the State Committee for Scientific
Research (Poland) under grant 1~P03B~078~26 in the period 2004-2006 and the
Grant-in-Aid for Scientific Research No.13135219 from the Japan Society
for the Promotion of Science.

\vskip 1.0cm
\centerline{APPENDIX}

\vspace*{0.3cm}
\noindent \hskip -0.46cm 
{\bf A1. Cross section for $\mib{\gamma\gamma\to t\bar{t}
\to \ell}^{\pm}\mib{X}$}

\noindent
For photons with definite polarizations the cross section for
$\gamma\gamma \to t\bar{t}$ is given by
\begin{eqnarray}
&&d\sigma(\gamma\gamma\to t\bar{t})
=C |{\cal M}_{\alpha\beta}
\epsilon^{\alpha}(h_1)\tilde{\epsilon}^{\beta}(h_2) |^2  \non\\
&&\phantom{d\sigma(\gamma\gamma)}
=C {\cal M}_{\alpha\beta}
\epsilon^{\alpha}(h_1)\tilde{\epsilon}^{\beta}(h_2)
{\cal M}_{\rho\sigma}^*
\epsilon^{*\rho}(h_1)\tilde{\epsilon}^{*\sigma}(h_2),
\end{eqnarray}
where $C$ expresses the kinematically-determined part, and $h_{1,2}$
are the helicities of the initial two photons:
\begin{equation}
\epsilon^{\mu}(h_1)
=\frac1{\sqrt{2}}(\epsilon^{\mu}_1 +ih_1 \epsilon^{\mu}_2),\ \ \
\tilde{\epsilon}^{\mu}(h_2)
=\frac1{\sqrt{2}}(-\epsilon^{\mu}_1 +ih_2 \epsilon^{\mu}_2)
\label{pol-vec}
\end{equation}
with $\epsilon^{\mu}_1=(0,1,0,0)$ and $\epsilon^{\mu}_2=(0,0,1,0)$.
Since $\epsilon^{\mu}(\pm 1)$ and $\tilde{\epsilon}^{\mu}(\pm 1)$ are
\[
\epsilon^{\mu}(\pm 1)
=\frac1{\sqrt{2}}(\epsilon^{\mu}_1 \pm i \epsilon^{\mu}_2),\ \ \
\tilde{\epsilon}^{\mu}(\pm 1)
=\frac1{\sqrt{2}}(-\epsilon^{\mu}_1 \pm i \epsilon^{\mu}_2),
\]
we can express $\epsilon^{\mu}(h_1)$ and $\tilde{\epsilon}^{\mu}(h_2)$
as
\begin{equation}
\epsilon^{\mu}(h_1)=\sum_{a=\pm 1} e^a \epsilon^{\mu}(a),\ \ \
\tilde{\epsilon}^{\mu}(h_2)
=\sum_{a=\pm 1} \tilde{e}^a \tilde{\epsilon}^{\mu}(a),
\end{equation}
where the coefficients are
$
e^{\pm 1} = (1 \pm h_1)/2$ and $\tilde{e}^{\pm 1} = (1 \pm h_2)/2$.
In terms of these quantities, the above cross section is
\begin{eqnarray}
&&d\sigma(\gamma\gamma\to t\bar{t})
=C \sum_{a,b,c,d=\pm 1} e^a e^{c*} \tilde{e}^b \tilde{e}^{d*}
{\cal M}_{\alpha\beta} {\cal M}_{\rho\sigma}^*
\epsilon^{\alpha}(a) \epsilon^{*\rho}(c)
\tilde{\epsilon}^{\beta}(b) \tilde{\epsilon}^{*\sigma}(d) \non \\
&&\phantom{d\sigma(\gamma\gamma\to t\bar{t})}
\equiv C \sum_{a,b,c,d=\pm 1} e^a e^{c*} \tilde{e}^b \tilde{e}^{d*}
{\cal T}^{ac,bd}\: .
\end{eqnarray}

%

The actual cross section $d\sigma(\gamma\gamma\to t\bar{t})$
measured at experiments is obtained by multiplying the above
cross section with the photon-spectra functions
$dN/dy$ and $dN/d\tilde{y}$, which work similarly to the
parton-distribution functions inside hadrons and replacing
$e^a e^{c*}$ and $\tilde{e}^b \tilde{e}^{d*}$ with the
spin density matrices $\rho$ and $\tilde{\rho}$ respectively,
%
\begin{equation}
d\sigma
=\sum_{a,b,c,d=\pm 1}\int dy\,d\tilde{y} \:
\frac{dN(y)}{dy} \frac{dN(\tilde{y})}{d\tilde{y}}
\rho^{ac}(y)\tilde{\rho}^{bd}(\tilde{y})
d\sigma^{\!\!ac,bd}(y,\tilde{y}), \label{cs1}
%
\end{equation}
where
\[
d\sigma^{ac,bd}(y,\tilde{y})=C\,{\cal T}^{ac,bd}(y,\tilde{y})
\]
is the cross section for two initial photons carrying
energy-fraction $y$ and $\tilde{y}$ of those of $e$ and $\bar{e}$,
$\rho(y)$ and $\tilde{\rho}(\tilde{y})$ are expressed in terms of
the three Stokes parameters as
\begin{equation}
\rho(y)
=\frac12\left(\begin{array}{cc}1+\xi_2(y) & \xi_3(y)-i\xi_1(y) \\
\xi_3(y)+i\xi_1(y) & 1-\xi_2(y) \end{array}\right),
\end{equation}
\begin{equation}
\tilde{\rho}(\tilde{y})=\frac12
\left(\begin{array}{cc}1+\tilde{\xi}_2(\tilde{y}) &
\tilde{\xi}_3(\tilde{y})+i\tilde{\xi}_1(\tilde{y}) \\
\tilde{\xi}_3(\tilde{y})-i\tilde{\xi}_1(\tilde{y}) &
1-\tilde{\xi}_2(\tilde{y}) \end{array}\right)
\end{equation}
in our choice of polarization vectors (\ref{pol-vec}).
Since $d\sigma^{ac,bd}(y,\tilde{y})$ is Lorentz-invariant, we can
calculate it in the $t\bar{t}(\gamma\gamma)$-CM frame.

The maximum of $y$ and $\tilde{y}$ is given by
\begin{equation}
y_{max}=\tilde{y}_{max}
=y_0/(1+y_0)\ \ {\rm and}\ \ 0 \leq y_0 \leq 2(1+\sqrt{2})
\simeq 4.828.
\end{equation}
Since we are interested in $t\bar{t}$ productions, $y$ and $\tilde{y}$
must satisfy
\begin{equation}
y\tilde{y} \geq 4m_t^2/s.
\end{equation}
Therefore the upper and lower bounds on these variables in the
integrations of eq.(\ref{cs1}) are
\begin{eqnarray}
&&y_{max}=y_0/(1+y_0),\ \ \ y_{min}=4m_t^2/(s \tilde{y}_{max}), \\
&&\tilde{y}_{max}=y_0/(1+y_0),\ \ \
  \tilde{y}_{min}=4m_t^2/(sy).
\end{eqnarray}

The photon-spectrum function $dN(y)/dy$ and $\xi_i(y)$ in the spin
density matrix $\rho(y)$ immediately after its production at the
conversion point are given by the following formulas:
\begin{eqnarray}
&&dN(y)/dy=C(y)/D(y_0), \\
&&\xi_1(y)=2P_t \sin(2\varphi) [r(y)]^2/C(y), \\
&&\xi_2(y)=[\:P_e f_2(y)+P_{\gamma} f_3(y)\:]/C(y), \\
&&\xi_3(y)=2P_t \cos(2\varphi) [r(y)]^2/C(y),
\end{eqnarray}
where
\begin{eqnarray}
&&C(y)=f_0(y)+P_e P_{\gamma}f_1(y), \\
&&D(y_0)=D_0(y_0)+P_e P_{\gamma}D_1(y_0), \\
&&f_0(y)=1/(1-y)+1-y-4r(y)[\:1-r(y)\:], \\
&&f_1(y)=y(2-y)[\:1-2r(y)\:]/(1-y), \\
&&f_2(y)=y_0 r(y)\bigl[\:1+(1-y)[\:1-2r(y)\:]^2\:\bigr], \\
&&f_3(y)=[\:1-2r(y)\:][\:1/(1-y)+1-y\:], \\
&&r(y)=y/[\:y_0(1-y)\:], \\
&&D_0(y_0)=(1-4/y_0 -8/y_0^2)\ln(1+y_0)+1/2 \non\\
&&\phantom{D_0(y_0)=}
 +8/y_0-1/[\:2(1+y_0)^2\:], \\
&&D_1(y_0)=(1+2/y_0)\ln(1+y_0)-5/2+1/(1+y_0) \non\\
&&\phantom{D_0(y_0)=}
 -1/[\:2(1+y_0)^2\:],
\end{eqnarray}
which are of course common to $dN(\tilde{y})/d\tilde{y}$ and
$\tilde{\xi}_i(\tilde{y})$ too.

Finally, combining thus-calculated $d\sigma(\gamma\gamma \to
t\bar{t})$ with $d{\mit\Gamma}(t \to {\ell}X)$ through the
Kawasaki-Shirafuji-Tsai formula \cite{technique}, we arrive
at the cross section $d\sigma(\gamma\gamma \to t\bar{t} \to
{\ell}^{\pm}X)$:
\begin{eqnarray}
&&\frac{d\sigma}{d\tilde{\mib{p}}_{\ell}} \equiv
\frac{d\sigma}{d\tilde{\mib{p}}_{\ell}}(1 + 2 \to t + \cdots \to \ell +
\cdots)  \non\\
&&\phantom{\frac{d\sigma}{d\tilde{\mib{p}}_{\ell}}}
=2\int d\tilde{\mib{p}}_t\frac{d\sigma}{d\tilde{\mib{p}}_t}(s_t = n)
\frac1{{\mit\Gamma}_t}\frac{d{\mit\Gamma}}{d\tilde{\mib{p}}_{\ell}}
=2B_{\ell}\int d\tilde{\mib{p}}_t
\frac{d\sigma}{d\tilde{\mib{p}}_t}(s_t = n)
\frac1{{\mit\Gamma}}\frac{d{\mit\Gamma}}{d\tilde{\mib{p}}_{\ell}}.
\label{KST1}
\end{eqnarray}
Here $d\tilde{\mib{p}}$ denotes the Lorentz-invariant phase-space
element $d\mib{p}/[\,(2\pi)^3 2p^0\,]$,
$d{\mit\Gamma}/d\tilde{\mib{p}}_{\ell}$ is the spin-averaged top-quark
width
\[
\frac{d{\mit\Gamma}}{d\tilde{\mib{p}}_{\ell}} \equiv
\frac{d{\mit\Gamma}}{d\tilde{\mib{p}}_{\ell}}(t\to \ell + \cdots),
\]
$B_{\ell}\equiv {\mit\Gamma}/{\mit\Gamma}_t$, and
$d\sigma(s_t = n)/d\tilde{\mib{p}}_t$ is the single-top-quark
inclusive cross section
\[
\frac{d\sigma}{d\tilde{\mib{p}}_t}(s_t = n) \equiv
\frac{d\sigma}{d\tilde{\mib{p}}_t}(1 + 2 \to t + \cdots \,;\:s_t=n)
\]
with the polarization vector $s_t$ being replaced with the so-called
``effective polarization vector" $n$
\begin{equation}
n_\mu = -\Bigl[\:g_{\mu\nu}-\frac{{p_t}_\mu{p_t}_\nu}{m_t^2}\:\Bigr]
{\sum_{\rm spin}\dps{\int}
d{\mit\Phi}\:\bar{B}{\mit\Lambda}_+\gamma_5 \gamma^\nu B \over
\sum_{\rm spin}\dps{\int}d{\mit\Phi}\:\bar{B}{\mit\Lambda}_+ B},
\label{n-vec}
\end{equation}
where the spinor $B$ is defined such that the matrix element for
$t(s_t)\to \ell +\cdots$ is expressed as $\bar{B}u_t(p_t,s_t)$,
${\mit\Lambda}_+\equiv \sla{p}_t +m_t$, $d{\mit\Phi}$ is the relevant
final-state phase-space element, and $\sum_{\rm spin}$ denotes the
appropriate spin summation. $d{\mit\Gamma}(t\to {\ell}X)$ for the
amplitude (\ref{ffdef}) is given by
\begin{equation}
\frac{1}{{\mit\Gamma}_t}
\frac{d{\mit\Gamma}}{dxd\omega}
=\frac{1+\beta}{\beta}\;\frac{3 B_{\ell}}{W}
\omega\Bigl[\:1+2{\rm Re}(f_2^R)\sqrt{r}
\Bigl(\frac{1}{1-\omega}-\frac{3}{1+2r} \Bigr)\:\Bigr]
\label{width}
\end{equation}
where $r \equiv (M_W/m_t)^2$, $W\equiv (1-r)^2(1+2r)$, 
$\omega \equiv (p_t -p_{\ell})^2/m_t^2$, and
$x$ is defined by the $\ttbar$ CM-frame lepton-energy
$E_{\ell}$ and $\beta\equiv\sqrt{1-4m_t^2/s}$ as
\[
x \equiv \frac{2E_{\ell}}{m_t}\sqrt{(1-\beta)/(1+\beta)}.
\]
Note that eq.(\ref{KST1}) holds very generally as long as
the narrow-width approximation is applicable to both the
top-quark and $W$-boson propagators.

\vskip 0.4cm
\noindent \hskip -0.46cm 
{\bf A2. Cross sections in $\mib{e}\bar{\mib{e}}$- and
$\mib{\gamma\gamma}$-CM frames}

\noindent
Calculations of cross sections are much simpler in $\gamma\gamma$-CM
frame, while actual experimental measurements are performed in $e\bar{e}
$-CM frame. Any observables expressed in these two frames are of course
connected via a proper Lorentz transformation. We summarize here some
necessary formulas.

Let us express the lepton energy and scattering angle in $e\bar{e}$-CM
frame as $E$ and $\theta$, and those in the $\gamma\gamma$-CM frame as
$E^{\star}$ and $\theta^{\star}$. Their relation is given by
\begin{equation}
E^{\star}=\frac{E(1-\beta\cos\theta)}{\sqrt{1-\beta^2}},\ \ \
\cos\theta^{\star}=\frac{\cos\theta-\beta}{1-\beta\cos\theta},
\label{Etheta-rel}
\end{equation}
where
\begin{equation}
\beta = (y-\tilde{y})/(y+\tilde{y}),
\end{equation}
$y$ and $\tilde{y}$ are the energy fractions of $e$ and $\bar{e}$
carried by the initial photon 1 and 2, which appeared in
eq.(\ref{cs1}). The scattering angle is defined as the angle
between the final-lepton $\ell$ and the initial-photon 1.

Based on these formulas, we have
\begin{eqnarray}
&&\frac{d\sigma}{dE d\cos\theta}
=\int dy\,d\tilde{y}
\frac{dN(y)}{dy} \frac{dN(\tilde{y})}{d\tilde{y}}
\frac{d\sigma}{dE d\cos\theta}(y,\tilde{y}) \non\\
&&\phantom{\frac{d\sigma}{dE d\cos\theta}}
=\int dy\,d\tilde{y}
\frac{dN(y)}{dy} \frac{dN(\tilde{y})}{d\tilde{y}}
J(y,\tilde{y})
\frac{d\sigma}{dE^{\star}d\cos\theta^{\star}}(y,\tilde{y}),
\label{cs-final}
\end{eqnarray}
where the Jacobian $J(y,\tilde{y})$ is given by
\begin{equation}
J(y,\tilde{y})=\frac{\sqrt{1-\beta^2}}{1-\beta\cos\theta}.
\end{equation}
$d\sigma(y,\tilde{y})/(dE^{\star}d\cos\theta^{\star})$ is
originally a function of $E^{\star}$ and $\cos\theta^{\star}$,
but they have to be re-expressed in terms of $E$ and
$\cos\theta$ in eq.(\ref{cs-final}) through the relations
given in eq.(\ref{Etheta-rel}).

Finally let us show the integration region for $E$ and $\cos
\theta$ in $d\sigma/(dEd\cos\theta)$. It is easy to confirm
\begin{equation}
-1 \leq \cos\theta^{\star} \leq +1\ \ \
\Longleftrightarrow\ \ \ -1 \leq \cos\theta \leq +1.
\end{equation}
On the other hand, the upper and lower bounds of the energy
are a bit more complicated. Since
\[
\frac{m_t r}{2}
\sqrt{
\smash{\frac{1-\beta^{\star}(y,\tilde{y})}{1+\beta^{\star}(y,\tilde{y})}}
\vphantom{A^2\over A}}
\leq
E^{\star}
\leq
\frac{m_t}{2}
\sqrt{
\smash{\frac{1+\beta^{\star}(y,\tilde{y})}{1-\beta^{\star}(y,\tilde{y})}}
\vphantom{A^2\over A}},
\]
where $\beta^{\star}(y,\tilde{y}) \equiv
\sqrt{1-4m_t^2/(y\tilde{y}s)}$,
one obtains
\begin{equation}
\frac12 m_t r
\sqrt{
\smash{\frac{1-\beta^{\star}(y,\tilde{y})}{1+\beta^{\star}(y,\tilde{y})}}
\vphantom{A^2\over A}}
\frac{\sqrt{1-\beta^2}}{1-\beta\cos\theta}
\leq E \leq
\frac12 m_t
\sqrt{
\smash{\frac{1+\beta^{\star}(y,\tilde{y})}{1-\beta^{\star}(y,\tilde{y})}}
\vphantom{A^2\over A}}
\frac{\sqrt{1-\beta^2}}{1-\beta\cos\theta}.
\end{equation}
Since the right(left)-hand side takes its maximum (minimum) for
$y=\tilde{y}=y_{max}$, we have
\begin{equation}
\frac12 m_t r 
\sqrt{\frac{1-\beta^{\star}_{max}}{1+\beta^{\star}_{max}}}
\leq E \leq
\frac12 m_t \sqrt{\frac{1+\beta^{\star}_{max}}{1-\beta^{\star}_{max}}}
\end{equation}
for the energy integration in the $e\bar{e}$-CM frame, where
$\beta^{\star}_{max}\equiv \beta^{\star}(y_{max},\tilde{y}_{max})$.

\vspace*{0.8cm}

\end{document}